\documentstyle[12pt]{article}

\begin{document}
\begin{titlepage}
\begin{flushright}\vbox{\begin{tabular}{c}
           TIFR/TH/96-53\\
           September, 1996\\
           hep-ph/yymmdd\\
\end{tabular}}\end{flushright}
\begin{center}
   {\large \bf
      $J/\psi$ Spin Asymmetries in the Colour-Octet Model.}
\end{center}
\bigskip
\begin{center}
   {Sourendu Gupta and Prakash Mathews.\\
    Theory Group, Tata Institute of Fundamental Research,\\
    Homi Bhabha Road, Bombay 400005, India.}
\end{center}
\bigskip
\begin{abstract}
We investigate the pattern of asymmetries in the colour-octet
model for inclusive production of charmonium with polarised
initial state hadrons and photons. We find that singly polarised
asymmetries vanish. The double polarised asymmetry for $J/\psi$
involves a new combination of long-distance matrix element. We
also investigate the asymmetries for $\eta_c$, $\chi_c$ and
$\Upsilon$ production. The asymmetries distinguish well between
different models of charmonium production.
\end{abstract}
\end{titlepage}

\section{Introduction}

A large volume of data on the inclusive hadro-production of $J/\psi$
has a good phenomenological description in terms of the non-relativistic
QCD (NRQCD) based colour octet model \cite{ours,br,our2}. There is no proof
of NRQCD factorisation for these cross sections and the agreement between
data and model may be considered phenomenological. Since other production
mechanisms for this cross section have also been postulated
\cite{baier,sld,twist}, it is useful to explore the pattern of spin
asymmetries in this model. Polarisation asymmetries for inclusive
charmonium production in various models have been investigated earlier
\cite{polcsm,polsld,anselmino}. Here we perform a detailed analysis in
the context of the colour-octet model. In the near future, experiments
at HERA and RHIC can use these predictions to assess the relative
importance of colour octet and other contributions to inclusive
$J/\psi$ production.

The physics of heavy quarks can be expected to be dominated by its (large)
mass, $m$, and hence computable in perturbative QCD. However, for quarkonium
systems the scales $mv$ (relative momentum) and $mv^2$ (energy) are also
important, since the quarks are expected to be non-relativistic
in the rest frame of the quarkonium. An organised way to include the effects
of the dimensionless parameter $v$ (relative velocity) is provided by an
effective low-energy field theory called non-relativistic QCD (NRQCD)
\cite{caswell}. A proof of factorisation can be given for several
processes \cite{bbl}.

In this framework, a heavy quark pair created by a short-distance process
binds into a quarkonium state over scales that are longer by powers
of $1/v$. Assuming factorisation, the inclusive cross-section for the
production of a meson $H$ can be written as
\begin{equation}
   d\sigma(H)\;=\;{1\over\Phi}d\Gamma
      \sum_{n=(\alpha,S,L,J)} {C(n)\over m^{d_n-4}}
       \left\langle{\cal O}^H_\alpha({}^{2S+1}L_J)\right\rangle
\label{e1}\end{equation}
where $\Phi$ is a flux factor, $d\Gamma$ is the phase space measure and the
$C$'s are short distance coefficients for the production of a heavy quark
pair of given colour ($\alpha$), spin ($S$) and orbital angular momentum
($L$) coupled to a total angular momentum ($J$). The naive scaling dimension
of the matrix element is $d_n$. It describes the long-distance
physics involved in the transition of the pair to the observed hadron
$H$. The colour octet model gets its name from the fact that the colour
index can run over singlet as well as octet representations.

The $C$'s have the usual perturbation expansion in powers of the strong
coupling $\alpha_{\scriptscriptstyle S}$. The matrix elements
$\langle{\cal O}^H_\alpha({}^{2S+1}L_J)\rangle$ scale as
$v^{3+2L+2E+4M}$ where $E$ and $M$ is the minimum number of gluon electric
and magnetic dipole transitions connecting the produced pair to the target
hadron. Consequently, the sum in eq.\ (\ref{e1}) is a double power series
expansion in $v$ and $\alpha_{\scriptscriptstyle S}$. In this paper we
consider all terms of order $\alpha_{\scriptscriptstyle S}^2v^n$ ($n\le7$)
for hadro-production and order $\alpha\alpha_{\scriptscriptstyle S}v^n$
($n\le7$) for (almost elastic) photo-production. Lepto-production will be
considered elsewhere.

Most of the matrix elements $\langle{\cal O}^H_\alpha({}^{2S+1}L_J)\rangle$
required for the spin-averaged $J/\psi$ cross section to this order are
now being extracted from data; some of them are known with fair accuracy.
They were first obtained from the phenomenology
of $P$-state charmonium production at large $p_{\scriptscriptstyle T}$
in Tevatron data \cite{jpsi}, where they are required for consistency
of the perturbation series \cite{bbl2}. The matrix elements for $S$-state
hadrons were also seen to be important for phenomenology \cite{brfl}.
A global analysis of Tevatron data has now been performed and many of
these matrix elements have been fixed \cite{cho1,sltk}. A linear combination
of octet matrix elements which appears for the first time in inclusive
$J/\psi$ production has been estimated through an analysis of (almost)
elastic photo-production data \cite{br,flem2}.

In section 2 we write down the coefficient functions appearing in
eq.\ (\ref{e1}) for the production of various resonances
with polarised initial states. The spin asymmetries are written down
in section 3. This section also contains a discussion of the kind of
the phenomenological signals expected from the colour octet model and
experimental checks on competing models. The concluding section contains
a discussion of the effects of higher orders in $v$.

\section{The Coefficient Functions}

In the past, the coefficient functions were computed using a covariant
projection technique \cite{baier,cho2,song}. A new algorithm for computing
these coefficient functions from the hard-scattering diagrams has recently
been developed \cite{bchen}. We follow this technique but use spherical
tensor methods for the simplification of angular momentum projections.
Our phase conventions are given in Appendix \ref{ap.tensor}.

The heavy quark ($Q$) and anti-quark ($\bar Q$) have
momenta $p$ and $\bar p$ respectively ($p^2=\bar p^2=m^2$). These are
written as
\begin{equation}
   p^\mu\;=\;{1\over2}P^\mu + L^\mu_i q^i\qquad{\rm and}\qquad
   \bar p^\mu\;=\;{1\over2}P^\mu - L^\mu_i q^i,
\label{momenta}\end{equation}
where $P$ is the centre of mass (CM) momentum of the pair and $q$ is the
relative momentum in the CM frame. This has only spatial (Euclidean) components
(denoted by Latin indices) and is boosted by the Lorentz transformation
$L^\mu_i$ with mixed Lorentz (in Greek) and spatial indices. The latter
contract over spatial indices of any space-like object and boosts it
properly, provided it has vanishing time component in the CM frame of the
pair.

In this paper we are concerned only with $2\to2$ processes giving
a $\bar QQ$ pair in the final state. If the initial 4-momenta are
taken to be $k_1$ and $k_2$, then $P=k_1+k_2$. Further, $K$ ($=k_1-k_2$)
is spacelike and orthogonal to $P$ and hence is also boosted properly
by $L^\mu_i$. Finally, for convenience, we choose the direction of
$K$ (in the rest frame of the pair) to be the $z$ direction and quantise
the angular momenta along this direction. In a
hadron-hadron collision this corresponds to choosing the common
direction of $k_1$, $k_2$ and $P$ to be the $z$ direction.

The perturbative QCD cross sections are written down as usual and then
matched to eq.\ (\ref{e1}) to extract the coefficient functions. In
order to do this, the Dirac spinors for $Q$ and $\bar Q$ are written in
terms of Pauli spinors ($\psi$ for $Q$ and $\chi$ for $\bar Q$), and the
matrix element is Taylor expanded in $q$. Since we only consider
terms upto order $v^7$, we can drop higher than linear
terms in $q$. The angular momentum projections are all effected by
recouplings of the spin-1 spherical tensor $L^\mu_i$. The matrix elements
for the processes $gg\to\bar QQ$ and $\bar qq\to\bar QQ$ have been
constructed and used to compute the coefficient functions for the cross
sections with unpolarised initial states. These provide checks on our
construction.

\subsection{$\bar qq\to\bar QQ$}
Keeping all terms upto linear order in $q$, we find that the colour
averaged squared matrix element for the lowest order $\bar qq$ process
is given by
\begin{equation}\begin{array}{rl}
   |{\cal A}|^2\;=\;&
     {\displaystyle4\pi^2\alpha_{\scriptscriptstyle S}^2\over\displaystyle9}
       (1-hh')\left(\delta_{ij}-\hat z_i\hat z_j\right)
     \langle\chi^\dagger\sigma^jT^a\psi\psi^\dagger\sigma^iT^a\chi\rangle,
\end{array}\label{qqmat}\end{equation}
where $h$ and $h'$ are the helicities of the two quarks, $\sigma$ denotes
a Pauli matrix and $T$ a colour generator. The terms linear in $h$ and $h'$
couple to the operator $i\epsilon_{ijk}\langle\chi^\dagger\sigma^i\psi
\psi^\dagger\sigma^j\chi\rangle$.
Since this is a negative parity operator, it does not appear in the NRQCD
action. As a result the single spin asymmetry vanishes. We discuss this
point later.

After inserting the proper projection operator, $\Pi_H(M)$, for the final
state hadron $H$ with helicity $M$, and converting Euclidean indices to
helicity indices ($\lambda$ and $\lambda'$), we obtain the expression
\begin{equation}\begin{array}{rl}
   |{\cal A}|^2\;=\;&(-1)^\lambda(1-hh')
     {\displaystyle \pi^2\alpha_{\scriptscriptstyle S}^2\over\displaystyle9m^2}
      \\&\quad\times
       \left[\delta_{\lambda,1}\delta_{\lambda',-1} +
             \delta_{\lambda,-1}\delta_{\lambda',1}\right]
     \left\langle\chi^\dagger\sigma^\lambda T^a\psi\Pi_H(M)
                  \psi^\dagger\sigma^{\lambda'}T^a\chi\right\rangle.
\end{array}\label{qqsht}\end{equation}
A summation over $\lambda$ and $\lambda'$ is understood. When
the helicity of the final state $H$ is observed, well-known angular momentum
techniques can be used to reduce the matrix
element to its standard form given in eq.\ (\ref{e1}). For the $J/\psi$, the
results of \cite{bchen} are reproduced. If the helicity is unobserved, then
a sum over $M$ can be performed, and the matrix element reduces to
$(-1)^\lambda\delta_{\lambda,-\lambda'}\langle{\cal O}^H_8({}^3S_1)\rangle/3$.

The last ingredients in the calculation of the coefficient functions are the
phase space measure and the flux factor--- 
\begin{equation}
   {d\Gamma\over\Phi}\;=\;{\pi\over4m^2}\delta(s-4m^2),
\label{phase}\end{equation}
where $s=P^2=(k_1+k_2)^2$.
The coefficient functions for the unpolarised cross section ($C$) and double
polarised asymmetries ($\tilde C$) can now be written down.
The results are---
\begin{equation}
    C({}^3S^{8}_1)\;=\;-\tilde C({}^3S^{8}_1)\;=\;
     {\displaystyle8\pi^2\alpha_{\scriptscriptstyle S}^2\over\displaystyle27m^2}.
\label{qqcof}\end{equation}
$C$ has been obtained previously in \cite{bchen}. Note that these coefficient
functions are to be multiplied by the dimension 3 operator matrix element
$\langle{\cal O}^H_{8}({}^3S_1)\rangle$. The parton level double polarised
asymmetry is $-1$ independent of the final hadron.

\subsection{$gg\to\bar QQ$}
We turn now to the $gg\to\bar QQ$ processes. In a ghost-free gauge there
are three of these--- an s-channel gluon exchange diagram and the $t$ and
$u$ channel heavy quark exchange diagrams. The summed matrix element breaks
up into three colour structures
\begin{equation}
   {\cal A}\;=\;-g^2 \epsilon^a_\mu(k_1,h) \epsilon^b_\nu(k_2,h')
     \left[{1\over6}\delta_{ab} S^{\mu\nu}+{1\over2}d^{abc}D^{\mu\nu}_c
           +{i\over2}f^{abc}F^{\mu\nu}_c\right].
\label{colour}\end{equation}
Note that Yang's theorem prevents the amplitudes $S$ and $D$ from
coupling to a spin-1 state \cite{yang}. No such restriction applies to the
amplitude $F$. Note also, that this theorem restricts only the state of the
$\bar QQ$ pair. The transition from this to the final observed hadron may
involve further soft gluon radiation. As a result, the angular momentum of
the final hadron is not restricted by Yang's theorem.

We take the density matrix for any initial state gluon of momentum
$k$ and helicity $h$ to be given by
\begin{equation}\begin{array}{rl}
   \epsilon^\mu(k;h)\epsilon^{*\nu}(k;h)\;=&\;
      {\displaystyle1\over\displaystyle2}
           \left[-g^{\mu\nu}+
         {\displaystyle k^\mu P^\nu+k^\nu P^\mu\over\displaystyle k\cdot P}
          -{\displaystyle P^2k^\mu k^\nu\over\displaystyle(k\cdot P)^2}\right]\\
     &\qquad\qquad
     -{\displaystyle ih\over\displaystyle2k\cdot P
                }\epsilon^{\mu\nu\alpha\beta}k_\alpha P_\beta.
\end{array}\label{gluon}\end{equation}
This gauge choice avoids ghosts due to the three gluon vertex.

The square of the amplitude $S$ after colour averaging is given by
\begin{equation}\begin{array}{rl}
   |S|^2 \;=&\;{\displaystyle\pi^2\alpha_{\scriptscriptstyle S}^2
                 \over\displaystyle36m^2}\biggl[(1+hh')\biggl\{
     \left\langle\chi^\dagger\psi\Pi_H(M)\psi^\dagger\chi\right\rangle
     +{\displaystyle2\over\displaystyle5m^2}
       \left(\delta_{\lambda,2}\delta_{\lambda',-2}+
               \delta_{\lambda,-2}\delta_{\lambda',2}\right)\\
       &\qquad\qquad\times
     \left\langle\chi^\dagger[\sigma q]^2_\lambda\psi\Pi_H(M)
                        \psi^\dagger[\sigma q]^2_{\lambda'}\chi\right\rangle
              \biggr\}\\&\qquad
     +{\displaystyle3\over\displaystyle m^2}(1-hh')
     \left\langle\chi^\dagger[\sigma q]^0_0\psi\Pi_H(M)
                        \psi^\dagger[\sigma q]^0_0\chi\right\rangle
         \biggr].
\end{array}\label{samp}\end{equation}
We have introduced the notation $[a b]^J_\lambda$ to denote two 3-vectors,
$a$ and $b$, coupled to a spherical tensor with angular momentum $J$ and
helicity $\lambda$. In this form the result is remarkably simple, compared
to the same expressions written out in Euclidean components. The square of
the $D$ amplitude is the same, apart from colour generators inserted between
the Pauli spinors. The square of the amplitude $F$ is simply
\begin{equation}\begin{array}{rl}
   |F|^2 \;=\;{\displaystyle3\pi^2\alpha_{\scriptscriptstyle S}^2
           \over\displaystyle32m^4}(1+hh')
     \left\langle\chi^\dagger q^0T^a\psi\Pi_H(M)
                  \psi^\dagger q^0T^a\chi\right\rangle,
\end{array}\label{famp}\end{equation}
with a helicity index on $q$.

In eq.\ (\ref{samp}) we have dropped terms linear in $h$ and $h'$, which
would contribute to the single spin asymmetries. These contain a non-zero
coefficient multiplying the operator
\begin{equation}
   \left\langle\chi^\dagger\psi\Pi(H)\psi^\dagger\sigma^i
        \left(-\,{i\over2}D^j\right)\chi\right\rangle + {\rm hc}.
\label{singlemat}\end{equation}
Since this is a negative parity operator, it does not appear in the NRQCD
action. As a result, single spin asymmetries vanish. The same argument
shows the absence of single spin asymmetries in $|D|^2$.

This mechanism is quite general. The single spin asymmetries are projected
by squaring the perturbative amplitude and contracting one set of indices
with the symmetric part of eq.\ (\ref{gluon}) and another with the
antisymmetric part. This would always couple to a negative parity operator.
Such operators cannot appear in the NRQCD action. Consequently single spin
asymmetries would vanish in all processes involving only strong or
electromagnetic vertices. Such an argument can also be developed for
$\bar qq$ initial states.

The $L=0$ ($S$-state) and $L=1$ ($P$-state) matrix elements have mass
dimensions 3 and 5 respectively. The coefficient functions can then be
written down---
\begin{equation}\begin{array}{rl}
   C({}^1S_0^{1})\;&=\;\tilde C({}^1S_0^{1})\;=\;
     {\displaystyle\pi^2\alpha_{\scriptscriptstyle S}^2\over\displaystyle9m^2},\\
   C({}^3P_0^{1})\;&=\;\tilde C({}^3P_0^{1})\;=\;
     {\displaystyle\pi^2\alpha_{\scriptscriptstyle S}^2\over\displaystyle3m^2},\\
   C({}^3P_2^{1})\;&=\;-\tilde C({}^3P_2^{1})\;=\;
     {\displaystyle4\pi^2\alpha_{\scriptscriptstyle S}^2\over\displaystyle9m^2},\\
   C({}^1S_0^{8})\;&=\;\tilde C({}^1S_0^{8})\;=\;
     {\displaystyle15\over\displaystyle8}C^0({}^1S_0^{1}),\\
   C({}^3P_0^{8})\;&=\;\tilde C({}^3P_0^{8})\;=\;
     {\displaystyle15\over\displaystyle8}C^0({}^3P_0^{1}),\\
   C({}^3P_2^{8})\;&=\;-\tilde C({}^3P_2^{8})\;=\;
     {\displaystyle15\over\displaystyle8}C^0({}^3P_2^{1}),\\
   C({}^1P_1^{8})\;&=\;\tilde C({}^1P_1^{8})\;=\;
     {\displaystyle\pi^2\alpha_{\scriptscriptstyle S}^2\over\displaystyle8m^2}.
\end{array}\label{form}\end{equation}
All the parton level asymmetries in $gg\to\bar QQ$ are exactly $1$ except the
${}^3P_2$ asymmetry, which is $-1$. This fact has interesting phenomenological
consequences.

\subsection{$\gamma g\to\bar QQ$}
The coefficient functions can be easily written down for photo-production
of $J/\psi$ in the almost elastic limit $z\approx1$. The inelasticity
parameter $z$ is defined by
\begin{equation}
   z\;=\;{p\cdot k_\psi\over p\cdot k_\gamma}\,,
\label{inelas}\end{equation}
where $p$, $k_\psi$ and $k_\gamma$ are the 4-momenta of the target proton,
final state charmonium and the photon respectively. The only amplitude
appearing in this process is proportional to the amplitude $D$ defined in
eq.\ (\ref{colour}). To linear order in $q$ there are three coefficient
functions--- $C({}^1S^8_0)$, $C({}^3P^8_0)$ and $C({}^3P^8_2)$. These differ
from the corresponding coefficients in eq.\ (\ref{form}) by an overall common
factor of $64\alpha/15\alpha_{\scriptscriptstyle S}$. Note that these
formul\ae{} can only be applied after subtracting the diffractive component
at $z=1$ \cite{diffr} and summing the remainder over a small range of $z$
(say $z>0.95$).

\section{Asymmetries}

In this section we consider the asymmetries for $J/\psi$ and other charmonium
states produced in polarised hadron-hadron or photon-hadron cross sections.
The quantities
\begin{equation}\begin{array}{rl}
   \hat\sigma\;=&\;{1\over4}\sum_{h,h'}\hat\sigma(h,h'),\\
   \Delta\hat\sigma\;=&\;{1\over4}\sum_{h,h'}hh'\hat\sigma(h,h'),
\end{array}\label{defs}\end{equation}
computed to order $v^7$ are listed in Appendix \ref{ap.cross}. Polarisation
asymmetries are given by
\begin{equation}
   \hat A\;=\;{\sum_{h,h'}hh'\hat\sigma(h,h')\over\sum_{h,h'}\hat\sigma(h,h')}
    \;=\;{\Delta\hat\sigma\over\hat\sigma}.
\label{asymdef}\end{equation}
Parton level cross sections and asymmetries are distinguished from
the physical (hadronic) quantities by a caret over any symbol for the
former.

The hadronic cross sections and asymmetries involve parton distributions in
the projectile (P) and target (T) hadrons. These are usually combined into
a parton luminosity factor which depends on whether one studies the integrated
cross section or distributions in the Feynman variable
$x_{\scriptscriptstyle F}$. For unpolarised initial states, the luminosities
relevant to $x_{\scriptscriptstyle F}$ distributions are---
\begin{equation}\begin{array}{rl}
   {\cal L}_g\;=&\; {\displaystyle1\over\displaystyle\sqrt{x_{\scriptscriptstyle F}^2+4\tau}}
       \,x_+g_{\scriptscriptstyle P}(x_+)
       \,x_-g_{\scriptscriptstyle T}(x_-),\\
   {\cal L}_q\;=&\; {\displaystyle1\over\displaystyle\sqrt{x_{\scriptscriptstyle F}^2+4\tau}}
       \left[\sum_f \,x_+q^f_{\scriptscriptstyle P}(x_+)
       \,x_-\bar q^f_{\scriptscriptstyle T}(x_-) + (P\leftrightarrow T)\right],
\end{array}\label{lumin}\end{equation}
where the sum in the definition of ${\cal L}_q$ is over flavours. Here $\tau=
4m^2/S$, where $S$ is the CM energy of the
colliding hadrons. These luminosities multiply the cross sections in eq.\ 
(\ref{cross}). We have used the notation
\begin{equation}
   x_\pm\;=\; {1\over2}(\sqrt{x_{\scriptscriptstyle F}^2+4\tau}\pm x_{\scriptscriptstyle F}).
\label{defbj}\end{equation}
The luminosities appropriate for total cross sections are given by integrals
over $x_{\scriptscriptstyle F}$. For polarised scatterings, the parton
distributions are replaced by the polarised parton distributions
$\Delta g(x)$ and $\Delta q(x)$. These polarised luminosities are denoted by
$\Delta{\cal L}_g$ and $\Delta{\cal L}_q$.

The hadronic cross sections in photo-production of charmonium involve
the polarised or unpolarised gluon densities at
\begin{equation}
   x\;=\;{2m^2\over m_{\scriptscriptstyle N} E_\gamma},
\label{xgm}\end{equation}
where $m_{\scriptscriptstyle N}$ is the nucleon mass and $E_\gamma$ is the
photon energy in the rest frame of the target nucleon.

\subsection{$\gamma p\to J/\psi$}

The polarisation asymmetry for photo-produced $J/\psi$ is given by
\begin{equation}
  A(\gamma p\to J/\psi)\;=\;
    \left({\Delta g(x)\over g(x)}\right)
    \left({\tilde\Theta\over\Theta}\right),
\label{jpsi.gm}\end{equation}
where the combination of matrix elements $\Theta$ and $\tilde\Theta$ are
given in eq.\ (\ref{spinsym}). Since $\tilde\Theta$ is unknown, it is not
possible to predict this asymmetry. Photo-production experiments must measure
$\tilde\Theta$. Present knowledge (summarised in Appendix \ref{ap.cross})
indicates that $A(\gamma p\to J/\psi)$ could be larger than
the gluon asymmetry $\Delta g/g$, and have the same sign as the latter.

Upto order $v^7$ $\chi_J$ states are not produced in $\gamma p$ reactions,
hence the asymmetries $A(\gamma p\to\chi_J)$ are undefined at this order.
For the same reason, $A(\gamma p\to\eta_c)$ is undefined at order $v^3$.

\subsection{$pp\to J/\psi$}

The colour octet model predictions for the polarisation asymmetries for
$\chi_0$, $\chi_2$ and $\eta_c$ are
\begin{equation}\begin{array}{rl}
  A(pp\to\eta_c)\;=\;&
    {\displaystyle\Delta{\cal L}_g\over\displaystyle{\cal L}_g},\\

  A(pp\to\chi_2)\;=\;&
    {\displaystyle
     -3\Delta{\cal L}_g\langle{\cal O}^{\chi_2}_1({}^3P_2)\rangle -
     10m^2\Delta{\cal L}_q\langle{\cal O}^{\chi_2}_8({}^3S_1)\rangle
     \over\displaystyle
     3{\cal L}_g\langle{\cal O}^{\chi_2}_1({}^3P_2)\rangle +
     10m^2{\cal L}_q\langle{\cal O}^{\chi_2}_8({}^3S_1)\rangle}\\

  A(pp\to\chi_0)\;=\;&
    {\displaystyle
     9\Delta{\cal L}_g\langle{\cal O}^{\chi_0}_1({}^3P_0)\rangle -
     8m^2\Delta{\cal L}_q\langle{\cal O}^{\chi_0}_8({}^3S_1)\rangle
     \over\displaystyle
     9{\cal L}_g\langle{\cal O}^{\chi_0}_1({}^3P_0)\rangle +
     8m^2{\cal L}_q\langle{\cal O}^{\chi_0}_8({}^3S_1)\rangle}\\
\end{array}\label{chi.p}\end{equation}
We can use the relations in eq.\ (\ref{hvq}) to simplify the results.
Since $\langle{\cal O}^{\chi_J}_1({}^3P_J)\rangle=0.32\pm0.04$ ${\rm
GeV}^5$ and $\langle{\cal O}^{\chi_1}_8({}^3S_1)\rangle=0.0098\pm0.0013$
${\rm GeV}^3$ \cite{cho1}, and ${\cal L}_g\gg{\cal L}_q$, the $\bar qq$
processes can be neglected to a good approximation. As a result we find
\begin{equation}
   A(pp\to\chi_0)\approx-A(pp\to\chi_2)
      \approx\Delta{\cal L}_g/{\cal L}_g=A(pp\to\eta_c).
\label{chi.pp}\end{equation}

The asymmetry for $J/\psi$ production can also be written down. Again, the
matrix elements for the $\bar qq$ subprocesses are much smaller than those
for the $gg$ subprocesses. As a result, to a good approximation, 
\begin{equation}\begin{array}{rl}
  A(pp\to J/\psi)\;=&\;
    \left({\displaystyle\Delta{\cal L}_g\over\displaystyle{\cal L}_g}\right)\\
      &\!\!\!\!\!\times
    {\displaystyle
     5m^2\tilde\Theta+15B_0\langle{\cal O}^{\chi_0}_1({}^3P_0)\rangle
              -4B_2\langle{\cal O}^{\chi_2}_1({}^3P_2)\rangle
     \over\displaystyle
     5m^2\Theta+15B_0\langle{\cal O}^{\chi_0}_1({}^3P_0)\rangle
              +4B_2\langle{\cal O}^{\chi_2}_1({}^3P_2)\rangle}
    +{\cal O}(2\%),
\end{array}\label{jpsi.pp}\end{equation}
where $B_J$ is the branching ratio for the decay of $\chi_J$ to $J/\psi$. If
$\tilde\Theta$ can be measured in $\gamma p$ reactions, then this asymmetry
is predicted without free parameters.

It should be understood that in eqs.\ (\ref{jpsi.gm}--\ref{jpsi.pp}),
the ratio of matrix elements is a kinematics-independent factor. Thus,
the same ratio appears for inclusive integrated asymmetries and for
$x_{\scriptscriptstyle F}$-dependence of the asymmetries. The kinematics
is subsumed into the parton density factors. As a result, apart from the
cross checks between hadro- and photo-production, each of these has
internal cross checks in the agreement between inclusive asymmetries and
their $x_{\scriptscriptstyle F}$-dependence.

\subsection{$\Upsilon$ production}
It is interesting to note the differences in $\Upsilon$ production
asymmetries. Since $v^2$ is much smaller for the bottomonium system than
for charmonium, and the branching fractions for $\chi_b\to\gamma\Upsilon$
are larger, the factors corresponding to $\tilde\Theta$ and $\Theta$ in
the analogue of eq.\ (\ref{jpsi.pp}) are relatively small. Consequently,
\begin{equation}
  A(pp\to \Upsilon)\;\approx\;
    \left({\displaystyle\Delta{\cal L}_g\over\displaystyle{\cal L}_g}\right)
    {\displaystyle
     15B_0\langle{\cal O}^{\chi_0}_1({}^3P_0)\rangle
              -4B_2\langle{\cal O}^{\chi_2}_1({}^3P_2)\rangle
     \over\displaystyle
     15B_0\langle{\cal O}^{\chi_0}_1({}^3P_0)\rangle
              +4B_2\langle{\cal O}^{\chi_2}_1({}^3P_2)\rangle}.
\label{ups.pp}\end{equation}
The colour octet components in this asymmetry are small, and the dominant
part is the same as in the colour singlet model.

In $\gamma p\to\Upsilon$ reactions, however, the colour octet parts (of
order $v^7$) are dominant, since there are no colour singlet parts at
lower order in $v$. Thus, the prediction for the polarisation asymmetry
is
\begin{equation}
  A(\gamma p\to J/\psi)\;=\;
    \left({\Delta g(x)\over g(x)}\right)
    \left({\tilde\Theta_b\over\Theta_b}\right),
\label{ups.gm}\end{equation}
where the subscript $b$ on the matrix elements denote that they are taken
for the bottomonium system.

\subsection{Distinction between models}

It is known that the colour octet contributions to the integrated $J/\psi$
cross section is important for unpolarised scattering. The normalisation of
the cross section requires the matrix element $\Theta$ defined in eq.\
(\ref{spinsym}). If these contributions were absent, as in the colour-singlet
model \cite{baier}, then, since $B_0$ is small,
\begin{equation}
  A(pp\to J/\psi)\;\approx\;A(pp\to\chi_2)
        \;=\;-{\displaystyle\Delta{\cal L}_g\over\displaystyle{\cal L}_g}
    \qquad({\rm colour\ singlet\ model}).
\label{singlet}\end{equation}
Falsification of this relation by the data would indicate the necessity of
colour octet components in the cross section. Detailed predictions based
on the colour singlet model have been extensively explored \cite{polcsm}.

An alternative model for the production of charmonium is the semi-local
duality model \cite{sld}. In this model the cross section for producing any
charmonium resonance is obtained by taking the sum of the $\bar qq\to\bar QQ$
and $gg\to\bar QQ$ cross sections (integrated in the range $4m^2\le S\le
4M_D^2$, where $M_D$ is the mass of the $D$ meson) and multiplying them by
a phenomenological number derived from the total cross sections. This model
is difficult to distinguish from the colour-singlet model using data on
unpolarised cross sections. However, the semi-local duality model predicts
\begin{equation}
  \begin{array}{rl}
  A(\gamma p\to H)\;=&\;{\displaystyle\Delta g(x)\over\displaystyle g(x)}\\
  A(pp\to H)\;=&\;
     {\displaystyle\Delta{\cal L}_g+\Delta{\cal L}_q
         \over\displaystyle{\cal L}_g+{\cal L}_q}
   \end{array}
     \qquad({\rm semi-local\ duality}),
\label{sld}\end{equation}
independent of the final state hadron $H$ \cite{polsld}. This is very
different from the prediction of the colour octet model, and an experimental
test should be simple.

Higher twist effects \cite{twist} give rise to more complicated
$x_{\scriptscriptstyle F}$ dependence of cross sections. Large effects of
this kind would violate the $x_{\scriptscriptstyle F}$ dependences
implicit in eqs.\ (\ref{jpsi.gm})--(\ref{ups.gm}).

\section{Discussion}

In section 3 we have considered all subprocesses upto order $v^7$. However,
from the analysis of data on the unpolarised cross sections, it is known that
certain aspects of the phenomenology require matrix elements upto order $v^9$
\cite{our2}. These will be considered in detail elsewhere. However, it is
possible to identify the major qualitative changes when these $v^9$ effects
are included.

The most important change is in the production of $\chi_J$. Upto order $v^7$
these resonances are produced only through the $\bar qq$ subprocess and the
amplitude $S$ in the $gg$ subprocess. As a result, upto this order, $A(\gamma
p\to\chi_J)$ is undefined. At order $v^9$, $\chi_J$ can be produced through
the $D$ amplitude in the $gg$ subprocess, for example by the matrix element
$\langle{\cal O}^{\chi_J}_8({}^1S_0)\rangle$. Intermediate pair states with
$L=2$ also occur. Consequently all these states can also be produced in
$\gamma p$ collisions. Therefore, $A(\gamma p\to\chi_J)$ becomes non-zero
and gets related to $\Delta g/g$.

At order $v^7$, $A(pp\to\chi_0,\chi_2)$ is related to $\Delta{\cal L}_g/
{\cal L}_g$, whereas $A(pp\to\chi_1)=\Delta{\cal L}_q/{\cal L}_q$. At order
$v^9$, because of the processes discussed above, $A(pp\to\chi_1)$ is well
approximated by $\Delta{\cal L}_g/{\cal L}_g$.

These changes in the $\chi_J$ cross sections and asymmetries reflect on
those for the $J/\psi$. Final state $J/\psi$'s can be produced by radiative
decays of these additional $\chi_J$'s. Eqs.\ (\ref{jpsi.gm}) and
(\ref{jpsi.pp}) will then involve new combinations of matrix elements.

Some of these considerations are very well illustrated by $\eta_c$ production
at order $v^7$. At the leading order $v^3$, $A(pp\to\eta_c)$ is given by
the expression in eq.\ (\ref{chi.p}). There is no $\eta_c$ production in
$\gamma p$ collisions at this order. At order $v^7$, however,
\begin{equation}
   \hat\sigma(\gamma g\to\eta_c)\;=\;
        \delta(s-4m^2){2\pi^3\alpha\alpha_{\scriptscriptstyle S}\over9m^3}
          \langle{\cal O}^{\eta_c}_8({}^1S_0)\rangle.
\label{chi.gm}\end{equation}
As a result, we obtain the asymmetry
\begin{equation}
   A(\gamma g\to\eta_c)\;=\;{\Delta g(x)\over g(x)}\,.
\end{equation}
In $pp$ collisions, order $v^7$ terms are obtained through the $D$ amplitude
which gives the same matrix element as in eq.\ (\ref{chi.gm}) above. In
addition, the matrix element $\langle{\cal O}^{\eta_c}_8({}^1P_1)\rangle$
enters through the $F$ amplitude in $gg\to\bar QQ$. The asymmetry becomes
\begin{equation}\begin{array}{rl}
   &A(\gamma g\to\eta_c)\;=\;
     {\displaystyle
      3\Delta{\cal L}_g\Psi-
      64{\cal L}_q\langle{\cal O}^{\eta_c}_8({}^3S_1)\rangle
     \over\displaystyle
      3\Delta{\cal L}_g\Psi+
      64{\cal L}_q\langle{\cal O}^{\eta_c}_8({}^3S_1)\rangle}\,,\\
   &\quad{\rm where\ }
    \Psi=8\langle{\cal O}^{\eta_c}_1({}^1S_0)\rangle +
         15\langle{\cal O}^{\eta_c}_8({}^1S_0)\rangle +
         {\displaystyle9\over\displaystyle m^2}
                \langle{\cal O}^{\eta_c}_8({}^1P_1)\rangle.
\end{array}\end{equation}
Since $\Psi$ is of order $v^3$ whereas $\langle{\cal O}^{\eta_c}_8({}^3S_0)
\rangle$ is of order $v^7$, one expects about 10\% corrections to the leading
order result (eq.\ \ref{chi.p}). This is a marginal effect, but may be
observable. Note that the linear combination $\Psi$ involves the matrix
element $\langle{\cal O}^{\eta_c}_8({}^1P_1)\rangle$, from the amplitude
$F$, not seen in $\eta_c$ production in $\gamma p$ collisions.

For the bottomonium system, the substantially smaller value of $v^2$
implies that these higher order corrections in $v$ are rather small.

The remaining part of the analysis presented in this paper is robust; the
results do not change as one takes higher orders in $v$ into account. We
conclude by summarising the most robust predictions of the colour octet model.
\begin{itemize}
\item
   All single spin asymmetries vanish in processes which do not involve
   at least one $W$ or $Z$.
\item
   Polarisation asymmetries for inclusive production and the
   $x_{\scriptscriptstyle F}$ distribution of any charmonium state
   involve the same combination of non-perturbative matrix element.
   The difference is only in the parton distribution factors.
\item
   Double spin asymmetries for $\chi_0$, $\chi_2$ and $\eta_c$ are
   simply related to the gluon asymmetries in hadro-production
   (eq.\ \ref{chi.pp}). The asymmetry for $\chi_2$ is (almost) equal
   and opposite to those for $\chi_0$ and $\eta_c$.
\item
   The $J/\psi$ asymmetries are simply related to gluon asymmetries
   in both hadro- and photo-production (eqs.\ \ref{jpsi.gm}, \ref{jpsi.pp}).
\item
   The charmonium asymmetries provide easy distinction between different
   models of charmonium production (eqs.\ \ref{singlet}, \ref{sld}).
\item
   The asymmetry in the hadro-production of $\Upsilon$ is very
   similiar to the results of the colour singlet model (eq.\ \ref{ups.pp}).
   That for photo-produced $\Upsilon$ is similiar to the $J/\psi$
   (eq.\ \ref{ups.gm}).
\end{itemize}

\appendix

\section{\label{ap.tensor}Phase Conventions for Spherical Tensors}

Conversion of Euclidean indices to helicity labels is effected by a unitary
transformation. This can be specified by the spherical components of a general
vector $V$---
\begin{equation}
   V_1\;=\;-{1\over\sqrt2}(V_x+iV_y),\qquad
   V_{-1}\;=\;-{1\over\sqrt2}(-V_x+iV_y),\qquad
   V_0\;=\;V_z.
\label{spherical}\end{equation}
With this phase convention, we may define the transformation of the Euclidean
metric---
\begin{equation}
   \delta_{ij}\;\to\;(-1)^\lambda\delta_{\lambda,-\lambda'}.
\label{metric}\end{equation}
Since the metric is not positive definite, it makes sense to
define covariant and contravariant indices. We choose the components of a
covariant vector to satisfy eq.\ (\ref{spherical}), and hence transform as
bras under rotations. The contravariant vector then transforms as a ket.

Clebsch-Gordan coefficients define unitary transformations between bases of
quantum states specified by two (commuting) angular momenta $j_1$ and $j_2$
and their resultant $J$, and is simply the matrix element
$\langle j_1,m_1;j_2,m_2|J,M\rangle$. Standard orthogonality relations
follow from the completeness of either basis. Following \cite{rose} we
choose these matrix elements to be real and fix $\langle j_1,j_1;j_2,j_2|
j_1+j_2,j_1+j_2\rangle=1$. Spherical tensors are irreducible components of
Euclidean tensors under rotations and can be constructed by first converting
Euclidean indices to helicity labels and then recoupling them through
the Clebsch-Gordan coefficients.

The metric can be extended to higher tensors, and is easily written in
terms of a Clebsch-Gordan coefficient---
\begin{equation}
   (-1)^\lambda\delta_{\lambda,-\lambda'}\;=\;
      \sqrt{(2J+1)}\langle J\lambda;J\lambda'|00\rangle.
\label{metcg}\end{equation}
The Levi-Civita tensor can also be expressed in terms of a Clebsch-Gordan
coefficient---
\begin{equation}
   \epsilon_{ijk}\;\to\;
      (-1)^\nu\langle1,\lambda;1,\mu'|1,-\nu\rangle.
\label{epscg}\end{equation}
These two tensors suffice for the problem at hand.

Simple angular momentum recouplings give the leading order heavy-quark
spin symmetry results
\begin{equation}\begin{array}{rl}
   \langle{\cal O}^{\chi_J}_1({}^3P_J)\rangle\;=&\;
      \langle{\cal O}^{\chi_{J'}}_1({}^3P_{J'})\rangle,\\
   \langle{\cal O}^\psi_{1,8}({}^3P_J)\rangle\;=&\;
      (2J+1)\langle{\cal O}^\psi_{1,8}({}^3P_0)\rangle,\\
   \langle{\cal O}^{\chi_J}_8({}^3S_1)\rangle\;=&\;
      (2J+1)\langle{\cal O}^{\chi_0}_1({}^3S_1)\rangle.
\end{array}\label{hvq}\end{equation}
Order $v^2$ corrections to these are harder to compute.

\section{\label{ap.cross}Cross Sections}

In this appendix we list the parton level cross sections and polarisation
asymmetries for charmonium production to order $v^3$ for $\eta_c$,
order $v^5$ for $\chi_J$ and order $v^7$ for $J/\psi$. The results for
unpolarised scattering have been obtained before \cite{cho1,bchen}.

For the $J/\psi$, the ${}^1S_0^{8}$ and the
${}^3P_J^{8}$ matrix elements are of order $v^7$. In addition, the
${}^3S_1^{8}$ matrix element from the $\bar qq$ process also contributes
to the same order. All other matrix elements are of higher order. However,
the $J/\psi$ final state can also be obtained by radiative decays of the
$\chi_J$ states. These arise at order $v^5$ through the ${}^3P_J^{1}$
matrix elements in the $gg$ subprocess and at the same order through the
${}^3S_1^{8}$ matrix element from the $\bar qq$ subprocess. Upto order
$\alpha_{\scriptscriptstyle S}^2v^7$ these are the only
channels which have to be summed to obtain the cross section for $J/\psi$
in hadro-production. For the $\eta_c$ the ${}^1S^1_0$ matrix element in the
$gg$ subprocess is the leading term. Other terms are subleading at least by
$v^4$.

The unpolarised cross sections are well-known \cite{baier,cho2,song,bchen}---
\begin{equation}\begin{array}{rl}
   \hat\sigma(gg\to J/\psi)\;=&\;\delta(s-4m^2)
     {\displaystyle\pi^3\alpha_{\scriptscriptstyle S}^2\over\displaystyle36m^3}
     \\&\!\!\!\!\!
     \left[ \left\langle{\cal O}^{J/\psi}_{8}({}^1S_0)\right\rangle
      +{\displaystyle3\over\displaystyle m^2}
         \left\langle{\cal O}^{J/\psi}_{8}({}^3P_0)\right\rangle
      +{\displaystyle4\over\displaystyle5m^2}
         \left\langle{\cal O}^{J/\psi}_{8}({}^3P_2)\right\rangle\right],\\

   \hat\sigma(gg\to \eta_c)\;=&\;\delta(s-4m^2)
     {\displaystyle\pi^3\alpha_{\scriptscriptstyle S}^2\over\displaystyle36m^3}
         \left\langle{\cal O}^{\eta_c}_{1}({}^1S_0)\right\rangle,\\

   \hat\sigma(gg\to \chi_0)\;=&\;\delta(s-4m^2)
     {\displaystyle\pi^3\alpha_{\scriptscriptstyle S}^2\over\displaystyle12m^5}
         \left\langle{\cal O}^{\chi_0}_{1}({}^3P_0)\right\rangle,\\

   \hat\sigma(gg\to \chi_2)\;=&\;\delta(s-4m^2)
     {\displaystyle\pi^3\alpha_{\scriptscriptstyle S}^2\over\displaystyle45m^5}
         \left\langle{\cal O}^{\chi_2}_{1}({}^3P_2)\right\rangle,\\

   \hat\sigma(\bar qq\to H)\;=&\;\delta(s-4m^2)
     {\displaystyle2\pi^3\alpha_{\scriptscriptstyle S}^2\over\displaystyle27m^3}
         \left\langle{\cal O}^H_{8}({}^3S_1)\right\rangle.
\end{array}\label{cross}\end{equation}
In the $\bar qq$ process the hadron $H$ can be taken to be anything. The
cross section for direct $J/\psi$ production involves the linear combination
of matrix elements determined in \cite{br,flem2}.

The polarisation asymmetries are given by $\Delta\hat\sigma/\hat\sigma$. The
numerators are listed here---
\begin{equation}\begin{array}{rl}
   \Delta\hat\sigma(gg\to J/\psi)\;=&\;\delta(s-4m^2)
     {\displaystyle\pi^3\alpha_{\scriptscriptstyle S}^2\over\displaystyle36m^3}
     \\&\!\!\!\!\!
     \left[ \left\langle{\cal O}^{J/\psi}_{8}({}^1S_0)\right\rangle
      +{\displaystyle3\over\displaystyle m^2}
         \left\langle{\cal O}^{J/\psi}_{8}({}^3P_0)\right\rangle
      -{\displaystyle4\over\displaystyle5m^2}
         \left\langle{\cal O}^{J/\psi}_{8}({}^3P_2)\right\rangle\right],\\

   \Delta\hat\sigma(gg\to \eta_c)\;=&\;\delta(s-4m^2)
     {\displaystyle\pi^3\alpha_{\scriptscriptstyle S}^2\over\displaystyle36m^3}
         \left\langle{\cal O}^{\eta_c}_{1}({}^1S_0)\right\rangle,\\

   \Delta\hat\sigma(gg\to \chi_0)\;=&\;\delta(s-4m^2)
     {\displaystyle\pi^3\alpha_{\scriptscriptstyle S}^2\over\displaystyle12m^5}
         \left\langle{\cal O}^{\chi_0}_{1}({}^3P_0)\right\rangle,\\

   \Delta\hat\sigma(gg\to \chi_2)\;=&\;-\delta(s-4m^2)
     {\displaystyle\pi^3\alpha_{\scriptscriptstyle S}^2\over\displaystyle45m^5}
         \left\langle{\cal O}^{\chi_2}_{1}({}^3P_2)\right\rangle,\\

   \Delta\hat\sigma(\bar qq\to H)\;=&\;-\delta(s-4m^2)
     {\displaystyle2\pi^3\alpha_{\scriptscriptstyle S}^2\over\displaystyle27m^3}
         \left\langle{\cal O}^H_{8}({}^3S_1)\right\rangle.
\end{array}\label{dcross}\end{equation}
The polarisation asymmetry for direct $J/\psi$ production contains a new
linear combination of matrix elements, as can be seen from the expression
above. The expressions for $\gamma g\to J/\psi$ cross sections can be obtained
by multiplying those for $gg\to J/\psi$ by
$64\alpha/15\alpha_{\scriptscriptstyle S}$.

Heavy quark spin symmetry can be used to reduce the matrix elements in
polarised and unpolarised direct $J/\psi$ production to the two
combinations---
\begin{equation}\begin{array}{rl}
    \Theta\;=\;
     \left\langle{\cal O}^{J/\psi}_{8}({}^1S_0)\right\rangle
      +{\displaystyle7\over\displaystyle m^2}
         \left\langle{\cal O}^{J/\psi}_{8}({}^3P_0)\right\rangle
    &\quad({\rm unpolarised}),\\
    \tilde\Theta\;=\;
     \left\langle{\cal O}^{J/\psi}_{8}({}^1S_0)\right\rangle
      -{\displaystyle1\over\displaystyle m^2}
         \left\langle{\cal O}^{J/\psi}_{8}({}^3P_0)\right\rangle
    &\quad({\rm polarised}).
\end{array}\label{spinsym}\end{equation}
There have been attempts to extract the combination $\Theta$ from
experiments; it is found that the value lies in the range $0.02$
and $0.03$ \cite{br,flem2}. Another combination,
\begin{equation}
   \vartheta\;=\;
     \left\langle{\cal O}^{J/\psi}_{8}({}^1S_0)\right\rangle
      +{\displaystyle3\over\displaystyle m^2}
         \left\langle{\cal O}^{J/\psi}_{8}({}^3P_0)\right\rangle,
\end{equation}
has also been extracted, and its value lies between $0.04$ and $0.066$
\cite{cho1,sltk}. The uncertainties reflect the spread in data as well
as the choice of parton densities, and perhaps also indicate the need
for higher order computations. If these two combinations are used to
determine $\tilde\Theta=2\vartheta-\Theta$, then we find that $\tilde
\Theta/\Theta$ lies between $1.7$ and $6.6$. This spread is larger than
the order $v^2$ errors involved in writing eq.\ (\ref{spinsym}).
In a previous work \cite{anselmino} it has been assumed, following a
suggestion in \cite{sltk}, that $\langle{\cal O}^{J/\psi}_8({}^3P_0)
\rangle\approx0$, and hence $\tilde\Theta/\Theta\approx1$.

%

\end{document}